\pdfoutput = 1
\documentclass[english, 11pt, a4paper]{article}
\usepackage{color}
\usepackage{bm}
\usepackage{amsmath}
\usepackage{wasysym}

\makeatletter

\usepackage{jcappub}
\usepackage{comment}

\makeatother

\usepackage{braket}

\usepackage{babel}
\begin{document}

\title{Probing Oscillons of Ultra-Light Axion-like Particle by 21cm Forest} 

\author[a,b]{Masahiro Kawasaki,}
\author[c]{Wakutaka Nakano,}
\author[a]{Hiromasa Nakatsuka,}
\author[a]{and Eisuke Sonomoto}

\affiliation[a]{ICRR, University of Tokyo, Kashiwa, 277-8582, Japan}
\affiliation[b]{Kavli IPMU (WPI), UTIAS, University of Tokyo, Kashiwa, 277-8583, Japan}
\affiliation[c]{Department of Physics, University of Tokyo, Tokyo, 113-0033, Japan}

\emailAdd{kawasaki@icrr.u-tokyo.ac.jp}
\emailAdd{m156077@icrr.u-tokyo.ac.jp}
\emailAdd{sonomoto@icrr.u-tokyo.ac.jp}
\emailAdd{nakatsuka@icrr.u-tokyo.ac.jp}

\abstract{
Ultra-Light Axion-like Particle (ULAP) is motivated 
as one of the solutions to the small scale problems in astrophysics.
When such a scalar particle oscillates 
with an $\mathcal{O}(1)$ amplitude in a potential shallower than quadratic,
it can form a localized dense object, oscillon.
Because of its longevity due to the approximate conservation of the adiabatic invariant,
it can survive up to the recent universe 
as redshift $z \sim \mathcal{O}(10)$.
The scale affected by these oscillons is determined by the ULAP mass $m$ 
and detectable by observations of 21cm line.
In this paper, 
we examine the possibility to detect ULAP by 21cm line and
find that the oscillon can enhance the signals of 21cm line observations when $m \lesssim 10^{-19}\ {\rm eV}$ and the fraction of ULAP to dark matter is much larger than $10^{-2}$
depending on the form of the potential.
}


\maketitle


\section{Introduction}

The nature of dark matter that dominates the matter energy density of the universe remains a huge mystery, 
though various observational results suggest its existence~\cite{Begeman:1991iy,Mateo:1998wg,Koopmans:2002qh}.
Thanks to the relentless efforts for cosmological observations over the past decades, 
particularly the observations of cosmic microwave background (CMB), 
cosmological constant and cold dark matter ($\Lambda$CDM) model 
with inflation is revealed to be the most promising~\cite{Hinshaw:2012aka,Abbott:2017wau,Aghanim:2018eyx}
among many cosmological models on large scales.

However, looking into the small scales around $\lesssim 1 {\rm Mpc}$,
numerical simulations based on the $\Lambda$CDM model confronts three astrophysical problems, 
missing satellite problem (e.g.~\cite{Moore:1999nt}), 
core cusp problem (e.g.~\cite{deBlok:2009sp}), 
and too big to fail problem (e.g.~\cite{BoylanKolchin:2011dk}) 
(see also Ref.~\cite{Bullock:2017xww} for a review).
Because all these problems arise from the over-density at small scales, 
scientists are struggling to construct the dark matter model that suppresses the small scale structure 
while behaves like cold dark matter at large scales.

Ultra-Light Axion-like Particle (ULAP) originated from the spontaneous symmetry breaking of string theory~\cite{Svrcek:2006yi}
is one of the fascinating particles that can solve such small scale problems. 
For example, considering the mass with $m\simeq10^{-22}\ {\rm eV}$, 
the de Broglie wavelength of ULAP is about $\sim$ kpc 
which is the typical scale of the galactic center. 
Smoothing out the central over density by the quantum pressure,
the core cusp problem can be solved~\cite{Hu:2000ke,Hui:2016ltb}.

Generally, ULAP is assumed to be coherently oscillating around the universe. 
However, there is a possibility that this scalar particle exists in the form of a localized dense object, 
oscillon~\cite{Bogolyubsky:1976yu,Gleiser:1993pt,Copeland:1995fq} (See \cite{Amin:2011hj,Amin:2010dc,Amin:2019ums} for earlier study of the formation.).
The necessary condition for oscillon formation is just the potential
shallower than quadratic. The lifetime of oscillons is estimated as
\begin{equation}
\tau \sim 2\ {\rm Gyr}\left(\frac{10^{-22}\ {\rm eV}}{m}\right)\left(\frac{\Gamma/m}{10^{-10}}\right)^{-1}.
\end{equation}
where $\Gamma$ is the decay rate of the oscillon 
which can be analytically calculated~\cite{Ibe:2019vyo,Zhang:2020bec}. 
Thus, if $\Gamma/m \lesssim 10^{-11}$, 
the produced oscillons can exist even in the current universe. 
The lifetime of oscillon is quite long in general 
because of the approximate conservation of the adiabatic charge~\cite{Kasuya:2002zs,Kawasaki:2015vga,Ibe:2019lzv}
while it also depends on the shape of the potential. 
In the pure natural type potential, for example, 
it is proved that the resultant oscillons are quite long-lived~\cite{Olle:2019kbo,Kawasaki:2019czd}. 
In this case, we can take advantage of the high density of oscillons to detect the clue to ULAP.

The oscillon formation affects fluctuations with comoving scale $m/\sqrt{10} \lesssim k/a \lesssim m$, 
that is,
\begin{align}
15\ {\rm Mpc^{-1}}\left(\frac{m}{10^{-22}\ {\rm eV}}\right)^{1/2} 
\lesssim k 
\lesssim 50\ {\rm Mpc^{-1}}\left(\frac{m}{10^{-22}\ {\rm eV}}\right)^{1/2},
\end{align}
where the upper bound is roughly determined by the typical distance between oscillons, 
and the lower bound is by the horizon scale at oscillon formation.
This is because the typical oscillon distance is the same as the wavenumber of the parametric resonance $\sim m$,
and oscillons are generally produced 
when the scale factor $a$ becomes $\mathcal{O}(10)$ times larger than  the initial value 
determined by the condition $H \simeq m$.

One of the methods to explore this scale is the 21cm line, 
which is produced by the hyperfine splitting by the interaction between the electron and proton spins~\cite{Madau:1996cs}. 
Generally, the 21cm line is adopted as the useful tracer of the recent billion years of the universe 
because neutral hydrogen is ubiquitous in the early universe after the recombination, 
amounting to $\sim75$\% of the gas present in the intergalactic medium (IGM). 

If there are luminous radio rich sources such as radio quasars
and gamma-ray bursts (GRBs), 
the emitted continuum spectrum
is consecutively absorbed by the neutral hydrogen; 
this absorption
mechanism is called 21cm forest~\cite{Carilli:2002ky, Furlanetto:2002ng} in analogy to the Lyman-$\alpha$ forest. 
The absorption could be the most efficient when the emission spectrum goes through the neutral hydrogen-rich region. 
Such regions during the epoch of reionization and beyond are called mini-halos 
characterized by the virial temperature smaller than $10^{4}\ {\rm K}$~\cite{Barkana:2000fd}.
Because under this temperature the metal-free cooling necessary for the star formation becomes ineffective and the amount of resultant X-rays is reduced,
plenty of neutral hydrogen remains in mini-halos.

In this paper, we focus on the detection of ULAP by 21cm forest when
some or all of ULAP is in the form of oscillon. 
In the previous researches~\cite{Shimabukuro:2019gzu,Shimabukuro:2020tbs},
the contribution of ULAP to the 21cm forest is discussed,
but the possibility of ULAP oscillon formation has never been considered.
Here, we assume that dark matter of the universe is consist of 
unknown cold dark matter, homogeneous ULAP, and ULAP oscillons,
\begin{align}
\Omega_{{\rm DM}} 
& =\Omega_{{\rm CDM}}+\Omega_{{\rm ULAP}},\\
& =\Omega_{{\rm CDM}}+\left(\Omega_{{\rm homo}}+\Omega_{{\rm osc}}\right).
\end{align}
For later use, 
we define the fraction of ULAP to cold dark matter $f_{{\rm ULAP}}$ as
\begin{equation}
\Omega_{{\rm ULAP}}\equiv f_{{\rm ULAP}}\Omega_{{\rm DM}}.
\end{equation}

The organization of this paper is as follows.
In Sec.~\ref{sec:Matter-Power-Spectrum}, 
we analytically derive the matter power spectrum under the situation 
where the ULAP oscillons are present in the universe following Ref.~\cite{Kawasaki:2020jnw}.
In Sec.~\ref{sec:absorption_abundance}, 
we calculate the abundance of the 21cm absorption lines.
Finally, in Sec.~\ref{sec:discussion} and Sec.~\ref{sec:conclusion} 
we discuss and conclude the result.
All cosmological parameters in this paper are extracted from the result of Planck 2018~\cite{Aghanim:2018eyx}.


\section{Matter Power Spectrum}
\label{sec:Matter-Power-Spectrum}

In this section, 
we briefly explain the matter power spectrum of the dark matter 
consist of unknown cold dark matter, homogeneous ULAP, and ULAP oscillons
following Ref.~\cite{Kawasaki:2020jnw}. 
The details of the derivation are written in Appendix \ref{sec:appnedix} and Ref.~\cite{Kawasaki:2020jnw}. 

The matter power spectrum $P(k)$ is decomposed as
\begin{align}
P(k,t) 
& = P_{{\rm CDM}}(k,t) + P_{{\rm ULAP}}(k,t),\\
& = P_{{\rm CDM}}(k,t) + \left[P_{{\rm homo}}(k,t) + P_{{\rm osc}}(k,t)\right].
\end{align}
where $P_{{\rm homo}}(k,t)$ and $P_{{\rm osc}}(k,t)$ show
the matter power spectra of homogeneous ULAP and ULAP oscillons, respectively 
and we assumed that the homogeneous part and the oscillon part are not correlated. 
We calculate $P_{{\rm {\rm CDM}}}(k,t)+P_{{\rm homo}}(k,t)$ from the AxionCAMB code~\cite{Hlozek:2014lca} 
which is originated from the public Boltzmann code CAMB~\cite{Lewis:1999bs,howlett2012cmb}.
Because homogeneous ULAP suppresses the small scale structure,
the matter power spectrum is also suppressed 
as $P_{\rm CDM}+P_{\rm homo} < P_{\Lambda{\rm CDM}}$,
which generally reduces the number of 21cm absoptions.

\subsection{ULAP Model}
\label{subsec:ULAP-Model}

As the ULAP potential, 
we choose the monodromy type potential~\cite{Silverstein:2008sg,McAllister:2008hb,Nomura:2017ehb}
\begin{equation}
V(\phi) = \frac{m^{2}F^{2}}{2p}\left[1-\left(1+\frac{\phi^{2}}{F^{2}}\right)^{-p}\right].
\label{eq:potential}
\end{equation}
When $p>-1$, the oscillon formation is confirmed in Ref.~\cite{Kawasaki:2019czd}.
It is also confirmed both analytically and numerically 
that the produced oscillons live very long~\cite{Olle:2019kbo,Ibe:2019vyo,Zhang:2020bec}.
In this paper, we take $p=-3/4$, for instance.

\subsection{Oscillon Matter Power Spectrum}

The analytical formula of the oscillon power spectrum has been developed in Ref~\cite{Kawasaki:2020jnw}
when the positions of produced oscillons are not correlated.
See Ref.~\cite{Kawasaki:2020jnw} and Appendix~\ref{sec:appnedix} for details of the derivation.
Defining the energy ratio of oscillons to ULAP $r_{\rm osc}(t)$ as
\begin{equation}
r_{\rm osc}(t)
\equiv \frac{\Omega_{\rm osc}}{\Omega_{\rm ULAP}},
\end{equation}
and neglecting the oscillon size ($k/a \ll m$),
the power spectrum at the oscillon formation time $t_f$ is written as 
\begin{equation}
  P_{\rm osc} \left(k, t_f \right)
  = 
  \frac{\left(r_{\rm osc}(t) f_{\rm ULAP}\right)^{2}}{n_{\rm osc}a^3} \left(\frac{\Omega_{\rm DM}}{\Omega_m}\right)^2
  \frac{\langle M_{\rm osc}^2 \rangle}{\langle M_{\rm osc} \rangle^2}
  \left[ 1 - \left( \frac{2}{kL_s} \right)^2 \sin^2\frac{kL_s}{2} \right],
  \label{eq:amplitude_estimation}
\end{equation}
where the bracket $\langle \rangle$ represents the ensemble average over oscillons, $M_{\rm osc}$ is the total energy of a oscillon,
and $n_{\rm osc}$ is the physical number density of oscillons.
$P_{\rm osc}$ is obtained by multiplying the Poisson power spectrum $1/(n_\text{osc} a^3)$ 
by the energy fraction of oscillons,
the squared average of the oscillon mass $\langle M_{\rm osc}^2 \rangle/\langle M_{\rm osc} \rangle^2 $,
and a suppression term.
The suppression is effective on scales larger than the horizon at the oscillon formation due to energy conservation.
Eq.~(\ref{eq:amplitude_estimation}) includes this suppression factor 
with $L_s$ being the cut-off scale~\cite{Kawasaki:2020jnw}.

Taking into account the time evolution after the oscillon formation,
this power spectrum is affected by two effects:
the partial decay of the oscillons
and the gravitational growth of the isocurvature fluctuations. 

\begin{figure}
  \centering{}
  \includegraphics[width=0.60\textwidth]{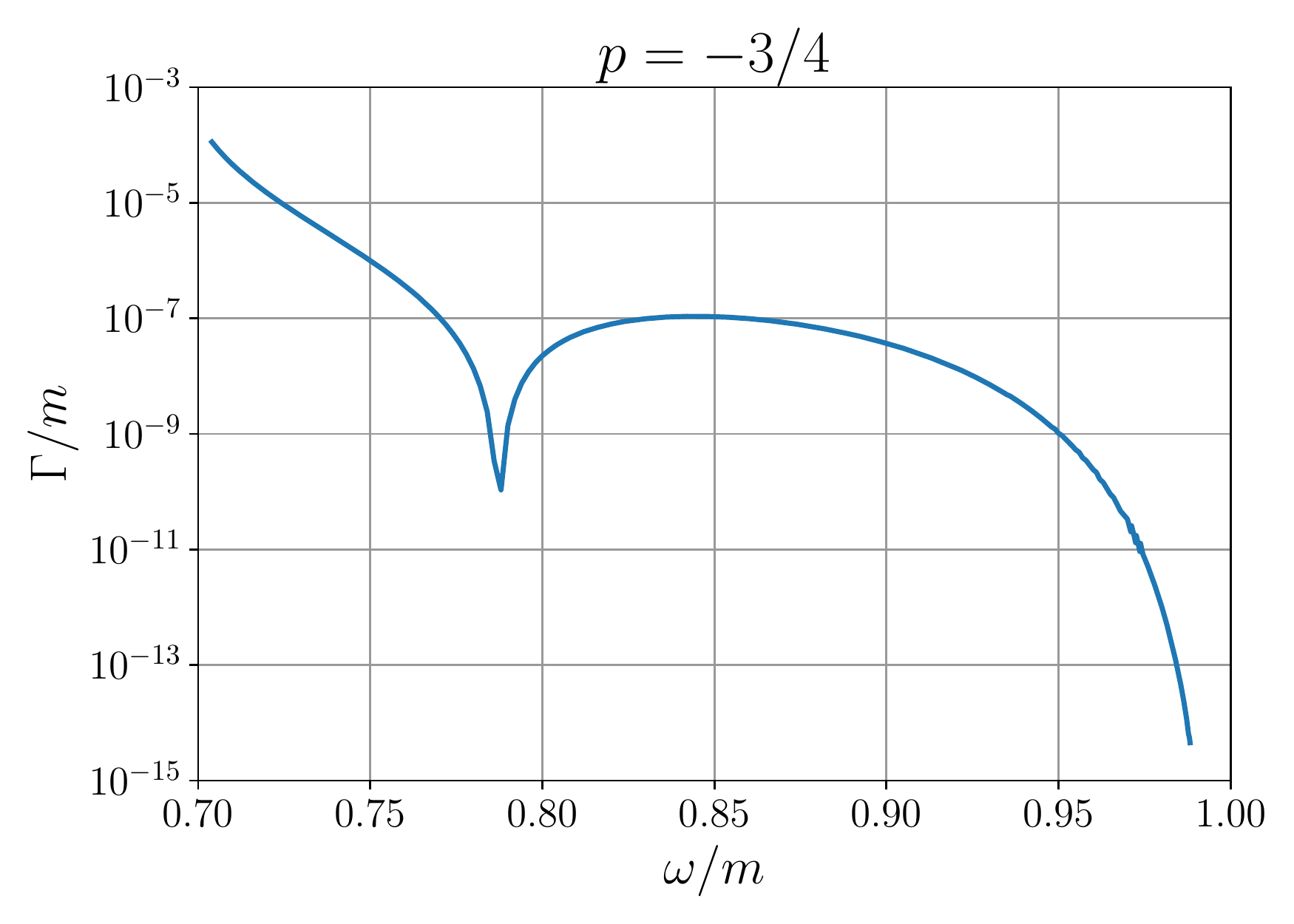}
  \caption{
    The decay rate of the oscillon for $p = -3/4$. 
    The horizontal axis $\omega$ shows the
    oscillation frequency of the scalar field.
    }
  \label{fig:decay_rate}
\end{figure}

First, let us consider the decay process of the produced oscillons. 
Because oscillons are getting smaller by emitting the self-radiation, 
the oscillon distribution also evolves.
Following Refs.~\cite{Ibe:2019vyo,Zhang:2020bec}, 
we can analytically calculate the oscillon decay rate $\Gamma$
\begin{equation}
  \Gamma\equiv\frac{1}{M_{\rm osc}}\left|\frac{dM_{\rm osc}}{dt}\right|.
\end{equation}
as shown in Fig.~\ref{fig:decay_rate}.
Using this decay rate,
we can evolve the oscillon distribution from the formation time.
The simulation result and the evolved distributions are shown in Fig~\ref{fig:histgram}. 
Please see Appendix~\ref{sec:setup} for the details of the simulation.

\begin{figure}[t]
  \centering{} 
  \includegraphics[width=0.9\textwidth]{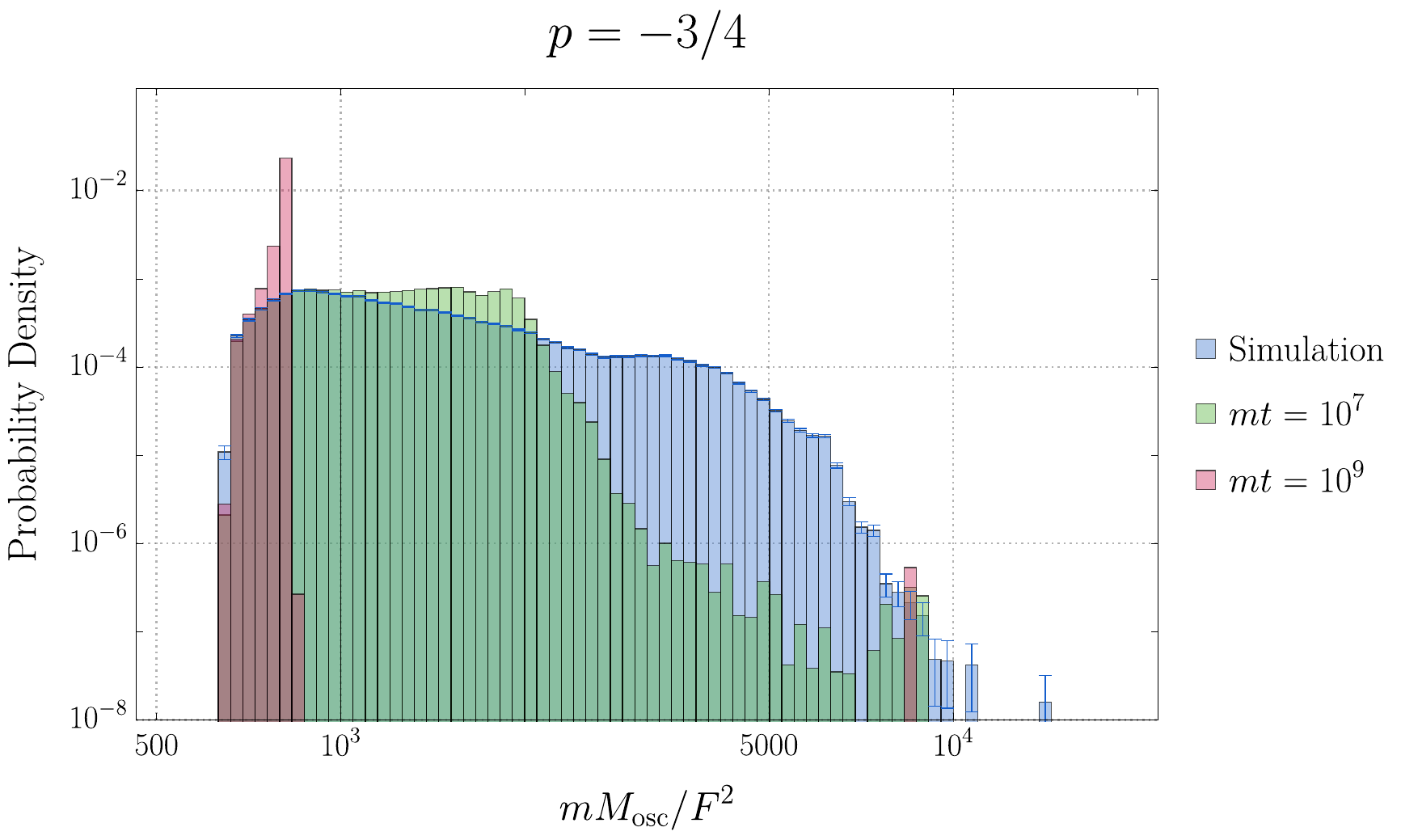}
  \caption{
    The probability density of the oscillon mass distribution derived from the lattice simulation
    which contains $\sim 10^5$ oscillons.
    The horizontal axis is normalized by the ULAP mass $m$ and the decay constant of ULAP $F$.
    The yellow region shows the oscillon distribution at the formation time,
    and the blue and green region show the distribution at $mt = 10^7$ and $mt = 10^9$, respectively.
    The width of the $i$-th bin is $\Delta x_i \equiv 10^{14/5} \left(10^{(i+1)/50}-10^{i/50} \right)$ 
    and statistical error of the simulation result is given by $\sqrt{N_i}/(N_{\rm osc}\Delta x_i)$ 
    where $N_i$ is the number of data in the $i$-th bin and $N_{\rm osc}$ is the total number of data.
  }
  \label{fig:histgram}
\end{figure}

\begin{figure}[ht]
  \centering{}
  \includegraphics[width=0.9\textwidth]{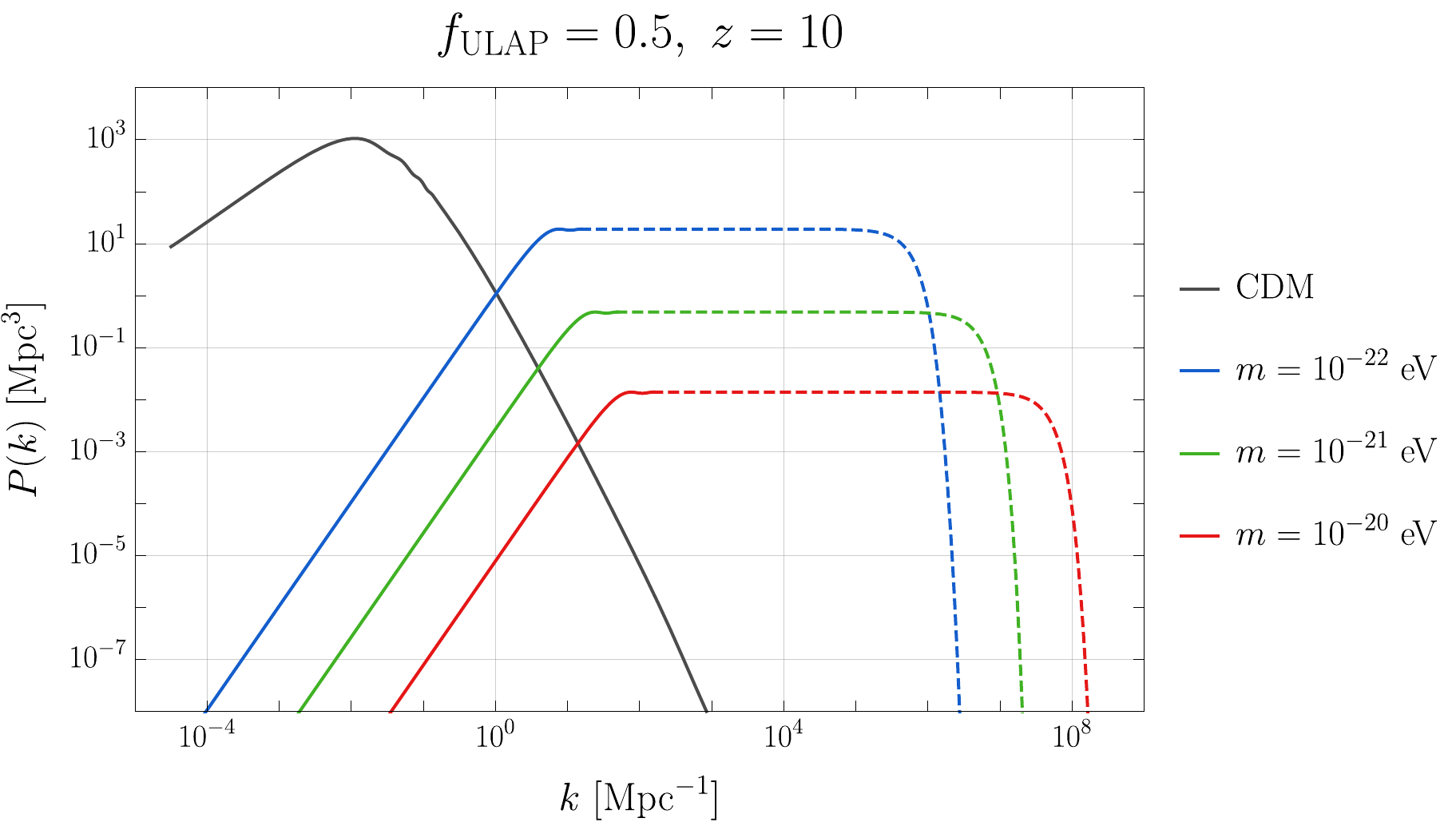}
  \caption{
    The matter power spectrum of oscillons when $f_{\rm ULAP} = 0.5$ at $z = 10$.
    The blue, green, and red lines show the oscillon matter power spectrum of $m = 10^{-22}\ {\rm eV},\ m = 10^{-21}\ {\rm eV},$ and $m = 10^{-20}\ {\rm eV}$ respectively.
    We truncated the matter power spectrum at the point where the oscillon number equals to $10$ 
    because the fluctuation of a single oscillon is non-linear. 
    These regions are plotted as dashed lines.
    Note that the exponential cutoff on the right side of the figure $k/a \sim m$
    is because of the radius of a oscillon~\cite{Kawasaki:2020jnw}.
    We also plotted the matter power spectrum of $\Lambda$CDM model as the black line for comparison.
    }
  \label{fig:power_sepctrum}
\end{figure}

The second is the growth of the fluctuations in the radiation and matter dominated era. 
In the parameter region of ULAP mass $m \gtrsim 10^{-28}\ {\rm eV}$ where we are interested,
oscillons are produced in the radiation dominated era. 
The fluctuations linearly grow after the matter-radiation equality
and the oscillon power spectrum in the matter dominated era ($t>t_{{\rm eq}}$) is
\begin{equation}
  P_{\rm osc}(k,t)
    = \left(\frac{3}{2}\frac{a}{a_{{\rm eq}}}\right)^{2} P_{\rm osc}(k, t_f).
\end{equation}
where $a_{\rm eq} \simeq 1/3400$ is the scale factor at the matter-radiation equality.

Considering these two effects, 
the oscillon matter power spectrum at $z = 10$ is calculated as shown in Fig.~\ref{fig:power_sepctrum}.
In the figure, we also take into account the non-linearity of the energy density of oscillons.
The linear matter power spectrum must be truncated at least below the scale 
where the oscillon number is smaller than $2$ 
because the fluctuation is non-linear in that scale. 
Thus, we cut off the power spectrum on the scale $k_{\rm cut}$ 
where the number of oscillons equals to $10$ as $n_{\rm osc} (2\pi/k_{\rm cut})^3 = 10$, for instance.
These lines are shown as dotted lines in Fig.~\ref{fig:power_sepctrum}.

\section{Abundance of 21cm Absorption Lines}
\label{sec:absorption_abundance}

In this section, 
we calculate the abundance of 21cm absorption lines 
when mini-halos contains ULAP oscillons.
The procedure of this section follows Refs.~\cite{Furlanetto:2002ng,Shimabukuro:2014ava,Shimabukuro:2019gzu,Shimabukuro:2020tbs}.

\subsection{Mini-Halo Profile}

To calculate the 21cm line absorption abundance, it is important to
estimate the abundance of the neutral hydrogen which absorbs the photon
of background light sources. In this subsection, we propose a decent
assumption of the matter distribution inside halos to derive the neutral
hydrogen distribution.

\subsubsection{Dark Matter Halo Profile}

The dark matter halo profile at the low redshift is well described
by the Navarro, Frenk, and White (NFW) profile~\cite{Navarro:1996gj,Hennawi:2005bm},
\footnote{
    We will discuss the validity of the NFW profile later in Sec.~\ref{sec:discussion}
}

\begin{equation}
\rho_{DM}(r)=\frac{\rho_{DM0}}{r/r_{s}\left(1+r/r_{s}\right)^{2}}\equiv\frac{\rho_{DM0}}{xy\left(1+xy\right)^{2}}.
\end{equation}
where $r_{s}$ is the scale papameter, $x,y$ are defined as $x\equiv r/r_{\rm vir},\ y\equiv r_{\rm vir}/r_{s}$, 
and $r_{\rm vir}$ is the virial radius. 
$y$ is often called the concentration paramter 
and fitted in Ref.~\cite{Comerford:2007xb}
\footnote{
    This fitting contains large uncertainty. 
    See also \cite{Bullock:1999he,Hennawi:2005bm}
}
as
\begin{equation}
   y = 
   \frac{14.8}{1+z}
   \left( \frac{M}{1.3 \times 10^{13}h^{-1}M_{\odot}} \right)^{-0.14},
\end{equation}
where we set $M$ as the virial mass, 
$M_{\odot}=1.33\times10^{33}\ {\rm kg}$ is the solar mass, 
and $h=0.68$ is the normalized Hubble parameter.

The virial radius $r_{\rm vir}$ is calculated by the spherical collapse model~\cite{Cooray:2002dia}, 
which derives
\begin{align}
  r_{\rm vir} 
  &= \left(\frac{M}{\frac{4\pi}{3}\rho_{m}(z)\Delta}\right)^{1/3},  \\
  &= 0.53\ {\rm kpc} \left(\frac{M}{10^8h^{-1} M_\odot}\right)^{1/3}
  \left(\frac{\Omega_{m0}}{\Omega_m}\frac{\Delta}{18\pi^2}\right)^{-1/3}\left(\frac{1+z}{10}\right)^{-1}.
\end{align}
where $\Delta\equiv18\pi^{2}+82d-39d^{2}$, and $d\equiv\Omega_{m}(z)-1$.
$\rho_{m}(z)$ is the matter energy density at the redshift $z$ 
and $\Omega_{m}(z)=\rho_{m}(z)/\rho_{c}(z)$ 
where $\rho_{c}(z)$ is the critical energy density at the redshift $z$.
From the above relations, 
the central energy density $\rho_{DM0}$ is determined as 
\begin{equation}
  \rho_{DM0}=\frac{M}{\int_{0}^{r_{\rm vir}}\left[r/r_{s}\left(1+r/r_{s}\right)^{2}\right]^{-1}d^{3}r}.
\end{equation}

\subsubsection{Gas Profile}

For simplicity, we make two assumptions on the gas within halos.
\begin{itemize}
\item Isothermal:
The gas within halos is isothermal because it is virialized. 
Defining the virial temperature as $\langle K\rangle = (3/2) k_B T_{\rm vir}$
where $\langle K\rangle$ is the time-averaged kinetic energy per particle of the system, the virial theorem leads to
\begin{align}
    T_{\rm vir} 
    \simeq \frac{\mu}{2k_B}\frac{GM}{r_{\rm vir}}.
\end{align}
Here $k_B$ is the Boltzmann constant 
and $\mu=1.22m_{p}$ is the mean molecular weight of the gas~\cite{Barkana:2000fd}.
\footnote{
    We ignore the $\mathcal{O}(1)$ coefficient of the gravitational potential.
}

\item In hydrostatic equilibrium:
At the distance $r$ from the origin, the gas pressure $P(r)$ and
the gravitational force are balanced~\cite{Makino:1997dv}.
\begin{equation}
\frac{dP(r)}{dr} = -\frac{GM(r)}{r^{2}}\rho_g(r),
\end{equation}
where $\rho_{g}(r)$ shows the gas profile. The gas pressure is easily
calculated from the equation of state as 
\begin{equation}
  P(r) = \frac{\rho_{g}(r)}{\mu} k_B T_{\rm vir}.
\end{equation}
From the above two assumptions, the gas profile $\rho_{g}(r)$ is
\begin{equation}
\rho_{g}(r)=\rho_{g0}\exp\left[-\frac{\mu}{2k_{B}T_{\rm vir}}\left(v_{{\rm esc}}(0)^{2}-v_{{\rm esc}}(r)^{2}\right)\right],
\end{equation}
where 
\begin{equation}
v_{{\rm esc}}(r)^{2}=2\int_{r}^{\infty}\frac{GM(\tilde{r})}{\tilde{r}^{2}}d\tilde{r}=\frac{2GM}{r_{\rm vir}}\frac{\log(1+xy)}{x\left[\log(1+y)-\frac{y}{1+y}\right]}.
\end{equation}
\end{itemize}

The normalization of the gas profile $\rho_{g0}$ is determined by
the ratio of the baryon energy density $\Omega_{b}$ and matter energy density $\Omega_{m}$ as
\begin{equation}
\frac{M_{g}}{M}=\frac{\Omega_{b}}{\Omega_{m}},\quad\Leftrightarrow\quad\rho_{g0}=\frac{\Delta}{3}\frac{y^{3}e^{A}}{\int_{0}^{y}(1+t)^{A/t}t^{2}dt}\frac{\Omega_{b}}{\Omega_{m}}\rho_{m}(z),
\end{equation}
where 
\begin{equation}
A\equiv\frac{3y}{\log(1+y)-y/(1+y)}.
\end{equation}

From all above calculations, the number deinsty profile of the neutral hydrogen is derived as
\begin{equation}
  n_{HI}(r) = 0.74 \times \frac{\rho_{g}(r)}{m_p}.
\end{equation}

\subsection{Halo Mass Function}
\begin{figure}[t]
  \centering{}
  \includegraphics[width=1.0\textwidth]{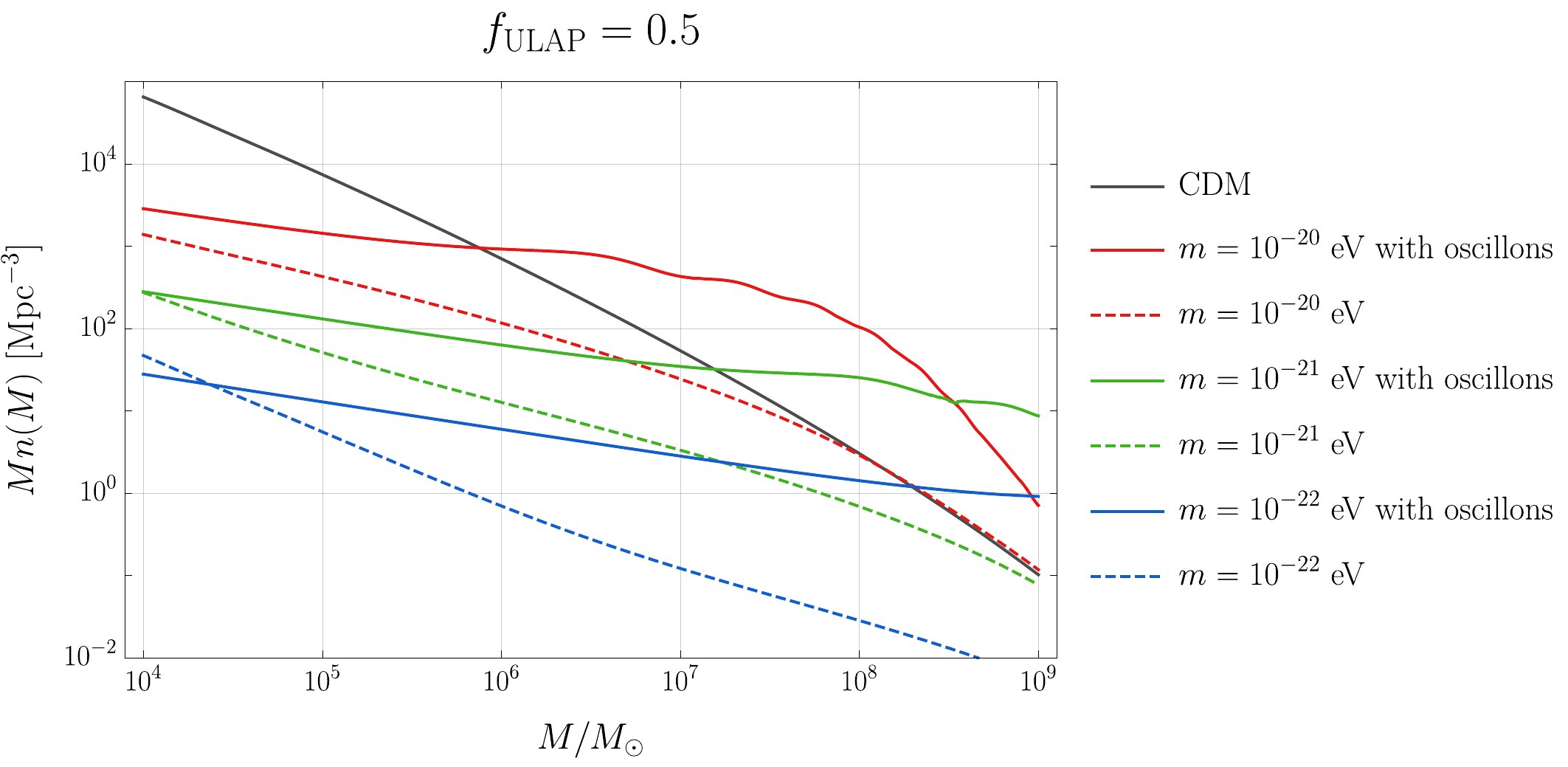}
  \caption{
      The halo-mass function when $f_{\rm ULAP} = 0.5$. 
      The red, green, and blue lines show the result of $m = 10^{-20}\ {\rm eV},\ m = 10^{-21}\ {\rm eV},$ and $m = 10^{-22}\ {\rm eV}$.
      The solid and dashed lines show the halo mass function with ULAP oscillons and without ULAP oscillon respectively.
      We also plotted the halo-mass function of $\Lambda$CDM model as the black line for comparison.
    }
  \label{fig:Press-Schechter}
\end{figure}
In this paper, 
we use the Press-Schechter formalism~\cite{Press:1973iz} to calculate the halo-mass function. 
Although the Sheth-Tormen mass function~\cite{Sheth:1999mn} is more precise in a low redshift, 
the situation is currently unclear for $z \sim 10$.
Thus, we use the Press-Schechter formalism in this work
and review it below.

First, we define the coarse-grained fluctuation as 
\begin{equation}
\delta_{{\rm cg}}(\bm{x},R)\equiv\int d^{3}x\delta(\bm{x})W(\bm{x},R),
\end{equation}
where we use the real space top-hat function as the window function $W(\bm{x},R)$
\footnote{
    The real space top-hat window function is described as
    \[
    W(\bm{x},R) 
    = \frac{3}{4\pi R^3} \Theta \left( R-|\bm{x}| \right).
    \]
    \[
    \therefore \quad  
    W(\bm{k},R) 
    = \frac{3}{(kR)^3} \left( \sin kR - kR \cos kR \right).
    \]
    where $\bm{x}$ is the comoving coordinate and $R$ is the comoving coarse-grained scale.
} 
in this paper. 
Assuming that the energy density fraction follows the Gaussian distribution, 
the probability that we can find the over-dense region where $\delta_{{\rm cg}}>\delta_{c}\simeq1.686$ in the linear perturbation theory is
\begin{equation}
P(>\delta_c)=\frac{1}{\sqrt{2\pi}\sigma(R)}\int_{\delta_{c}}^{\infty}\exp\left(-\frac{\delta_{{\rm cg}}^{2}}{2\sigma(R)^{2}}\right)d\delta_{cg},
\end{equation}
where $\sigma(R)$ is the coarse-grained variance and
\begin{align}
\sigma(R)^{2} & \equiv\langle\delta_{cg}^{2}(\bm{x},R)\rangle=\int P(k)\left|W(kR)\right|^{2}d^{3}k.
\end{align}
We substitute the matter power spectrum derived in Sec.~\ref{sec:Matter-Power-Spectrum} for $P(k)$. 
Because the coarse-grained comoving radius $R$ is related to the halo mass $M$ as
\begin{equation}
M = \frac{4\pi R^{3}}{3}\rho_{m0},
\end{equation}
we can take $\sigma$ as the function of $M$ instead of $R$. 
Then, the comoving number density of halos with the mass $M$ is calculated from the Press-Schechter formalism as
\begin{align}
n(M,z) dM
&= 2 \frac{\rho_{m0}}{M} \frac{\partial P(>\delta_{c})}{\partial M}dM,\\
&= \sqrt{\frac{2}{\pi}}\frac{\rho_{m0}}{M}\frac{\delta_{c}}{\sigma(M)^{2}}\exp\left(-\frac{\delta_{c}^{2}}{2\sigma(M)^{2}}\right)\left|\frac{\partial\sigma(M)}{\partial M}\right|dM.
\end{align}

The calculation results are shown in Fig.~\ref{fig:Press-Schechter}.
When ULAP without oscillons exists as dark matter,
the number of halos is suppressed compared to the CDM case
because of the quantum pressure of ULAP as shown in dashed lines.
On the other hand, the solid lines which exhibit the halo-mass function with ULAP oscillons are always larger than the corresponding dashed lines, 
because of the amplification of the matter power spectrum by ULAP oscillons.
They are even larger than that of the CDM case depending on the ULAP mass.

\subsection{Spin Temperature}

The common way to describe the number density ratio between the excited
state $n_{1}$ and the ground state $n_{0}$ is the spin temperature
$T_{s}$ \cite{4065250}. Assuming that the neutral hydrogen follows
the Boltzmann distribution, the spin temperature is defined as
\begin{align}
\frac{n_{1}}{n_{0}} & \equiv\frac{g_{1}}{g_{0}}\exp\left(-\frac{E_{10}}{k_{B}T_{s}}\right)\equiv3\exp\left(-\frac{T_{*}}{T_{s}}\right),
\end{align}
where $g_{1}=3,\ g_{0}=1$ are the degrees of freedom of the excited
state and the ground state respectively, and $T_{*}\equiv E_{10}/k_{B}\simeq68\ {\rm mK}$.

In the two-level system of neutral hydrogen, there are three processes we should take
into account, spontaneous emission, excitation, and stimulated emission.
The rate of the spontaneous emission is calculated as $A_{10}=2.85\times10^{-15}\ {\rm s}^{-1}$
and the rate of the other two processes depends on the details
of the three interactions below.
\begin{itemize}
\item CMB

Neutral hydrogens are excited or deexcited by the interaction with
the CMB photons. Here, we define the transition rates for the excitation and stimulated
emission as $B_{01}(\nu)\rho_{{\rm CMB}}(\nu)$
and $B_{10}(\nu)\rho_{{\rm CMB}}(\nu)$, respectively.
Here, $\rho_{{\rm CMB}}(\nu)$ obeys the Planck
distribution.
\item Collisions

Collisions of neutral hydrogens affect the spin temperature by
swapping the electron spin. There are two main processes in this collision
interaction: $H-H$ and $H-e$ collisions. Suppose that
the deexcitation rate is defined as $C_{10}$, the transition rate
is decomposed into two parts,

\begin{equation}
C_{10}=n_{HI}\kappa_{10}^{HH}+n_{e}\kappa_{10}^{eH},
\end{equation}
where $\kappa_{10}^{HH},\ \kappa_{10}^{eH}$ are the deexcitation
rates of $H-H$ and  $H-e$ collisions respectively~\cite{Zygelman_2005,Furlanetto:2006jb}
and $n_{HI},\ n_{e}$ are the number densities of the neutral hydrogen
and electron respectively. Because we consider the universe where
the reionization is not still effective, the fraction of free electrons
is small and we can safely ignore the effect of the $H-e$ collision. Thus,
we assume that almost all gases are consist of neutral hydrogen,
that is, $C_{10}\simeq n_{HI}\kappa_{10}^{HH}$. We also define the
excitation rate as $C_{01}$. 
\item Lyman-$\alpha$ photons

Lyman-$\alpha$ photons also affect the state of the neutral hydrogen
by the transition via Lyman-$\alpha$ energy level,
called Wouthuysen--Field effect. 
Let us define the excitation and deexcitation rates as $P_{01}$ and $ P_{10}$, respectively. The Lyman-$\alpha$
photons are mainly created from stars, but the star formation process within $z \lesssim 30$ strongly depends on the astrophysics and contains
uncertainties. Thus, we ignore the Lyman-$\alpha$ contribution for just
simplicity in this paper.
\end{itemize}

In a similar way to the spin temperature, 
we introduce the gas kinetic temperature $T_K$ and the color temperature $T_c$ as
\begin{align}
\frac{C_{01}}{C_{10}} & =3\exp\left(-\frac{T_{*}}{T_{K}}\right),\quad\frac{P_{01}}{P_{10}}=3\exp\left(-\frac{T_{*}}{T_{c}}\right).
\end{align}
When all these three processes are in equilibrium, assuming $T_{*}\ll T_{K},T_{C},T_{\gamma}$
the spin temperature is written as

\begin{equation}
T_{S}^{-1}\simeq\frac{T_{\gamma}^{-1}+x_{c}T_{K}^{-1}+x_{\alpha}T_{\alpha}^{-1}}{1+x_{c}+x_{\alpha}}\simeq\frac{T_{\gamma}^{-1}+x_{c}T_{\rm vir}^{-1}}{1+x_{c}},
\end{equation}
where
\begin{equation}
x_{c}\equiv\frac{C_{10}}{A_{10}}\frac{T_{*}}{T_{\gamma}},\quad x_{\alpha}\equiv\frac{P_{10}}{A_{10}}\frac{T_{*}}{T_{\gamma}},
\end{equation}
and $T_{\gamma}(z)=2.73(1+z)\ {\rm K}$ is the CMB photon temperature.
We have used the fact that the gas temperature $T_{K}$ is well described
by the virial temperature $T_{\rm vir}$ when the reionization is not
complete. As we mentioned, we ignore the contribution from Lyman-$\alpha$
photons, that is, $x_{\alpha}=0$.

\subsection{Optical Depth}

The optical depth is obtained by the integration of the absorption
coefficient 
over the entire distance as
\begin{align}
\tau(\nu,M,\alpha) 
& =\frac{3hc^{2}A_{10}}{32\pi k_{B}\nu_{10}}\int_{-R_{max}}^{R_{max}}\frac{n_{HI}(r)}{T_S(r)}\phi(\nu)dR,
\end{align}
where $\alpha$ is the impact parameter,
$h$ is Planck constant, $\nu_{10}=1420\ {\rm MHz}$,
$c$ is the speed of light,
$R = \sqrt{r^2 - \alpha^2}$ is the coordinate along the line of sight,
and
\begin{equation}
\phi(\nu)=\frac{c}{\sqrt{\pi}b}\exp\left[-\frac{c^{2}\left(\nu^{2}/\nu_{10}^{2}-1\right)}{b^{2}}\right],\quad\left(b\equiv\frac{2k_{B}T_{\rm vir}}{m_{p}}\right)
\end{equation}
is the line profile function. Here we only consider the Doppler broadening effect due to the thermal dispersion of the neutral hydrogen. 

\subsection{Abundance of 21cm Absorbers}
\begin{figure}[t]
\centering{}
\includegraphics[width=1.0\textwidth]{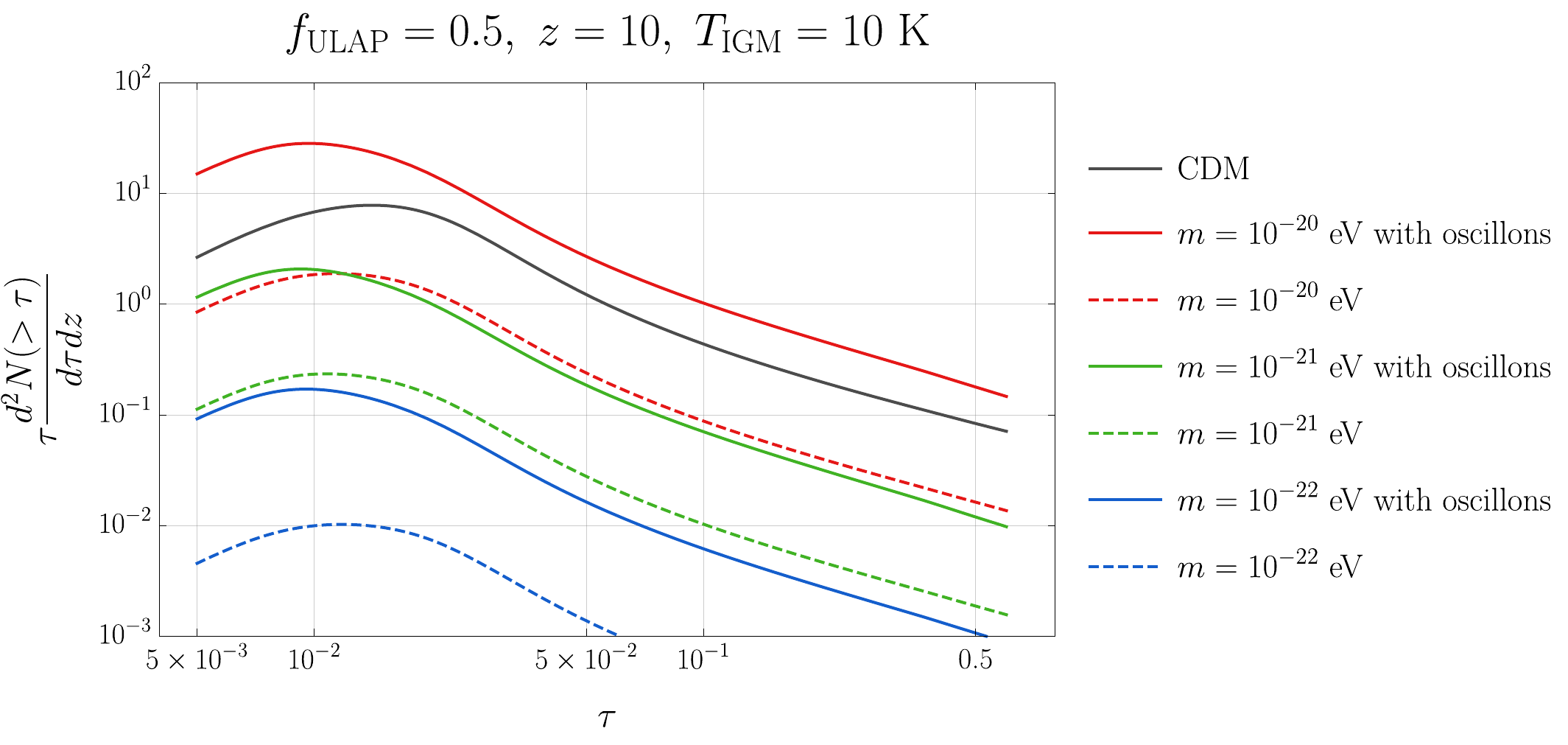}
\includegraphics[width=1.0\textwidth]{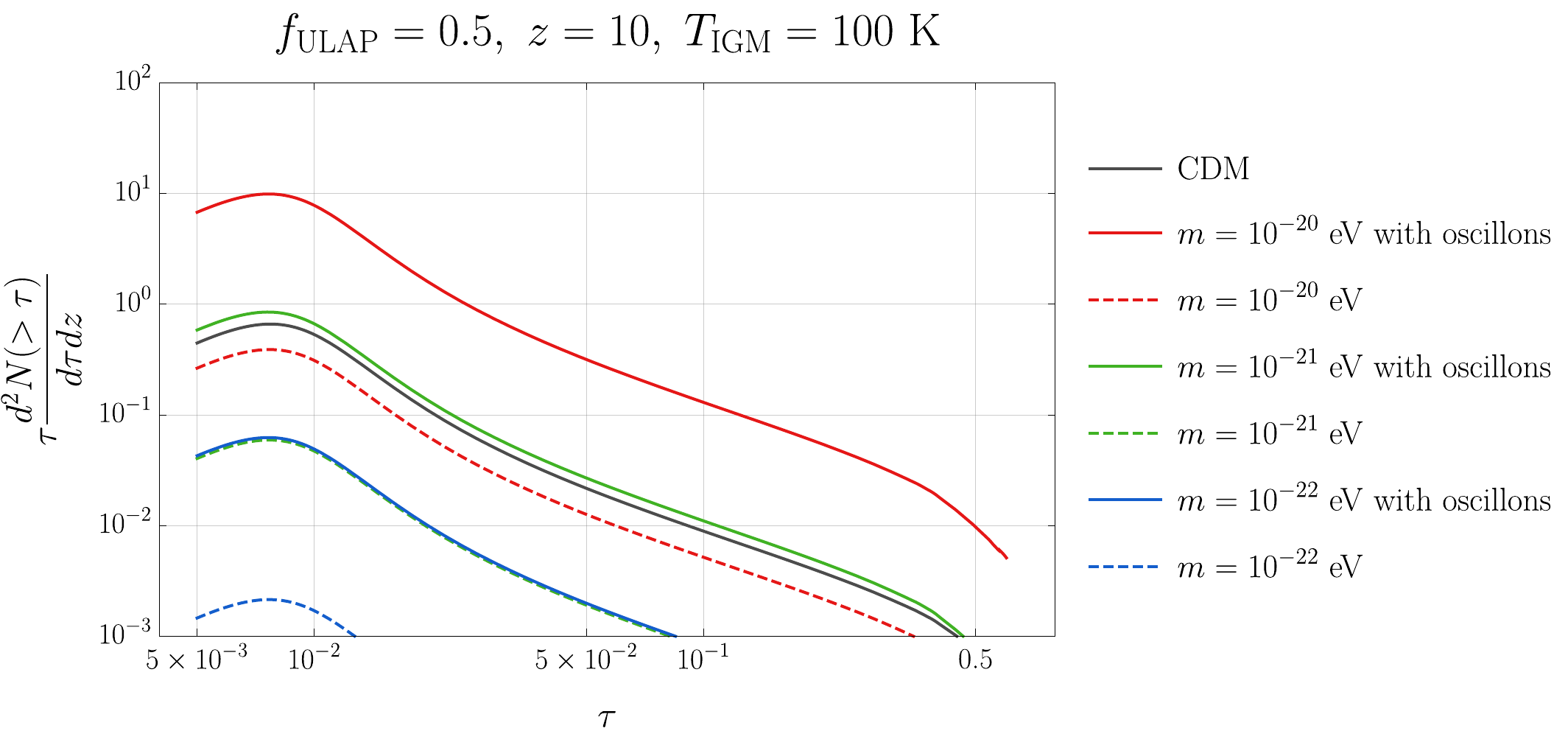}
\caption{
    The adundance of 21cm absorbers.
    The upper and lower figures show the case of $T_{\rm IGM} = 10\ {\rm K}, 100\ {\rm K}$ respectively.
    The horizontal axis shows the optical depth 
    and the vertical axis does abundance of the number of systems intersected with the optical depth $\tau$ per redshift interval.
    The red, green, and blue lines show the result of $m = 10^{-20}\ {\rm eV},\ m = 10^{-21}\ {\rm eV},$ and $m = 10^{-22}\ {\rm eV}$.
    and the solid and dashed lines show the halo mass function with ULAP oscillons and without ULAP oscillon respectively.
    We also plotted the result of $\Lambda$CDM model as the black line for comparison.
}
\label{fig:absorbers}
\end{figure}
From the above relations, 
we can derive the number of systems intersected with the optical depth greater than $\tau$ per redshift interval as
\begin{equation}
    \frac{dN(>\tau)}{dz}
    = (1+z)^2 \frac{dr}{dz} \int_{M_{\rm min}}^{M_{\rm max}} n(M,z) \pi \alpha(\tau)^2 dM,
\end{equation}
where $dr/dz=c/H(z)$ is the comoving line element 
and $\alpha(\tau)$ shows the maximum physical radius where the optical depth exceeds $\tau$. 
As the halo mass function, 
we will use the matter power spectrum derived in Sec.~\ref{sec:Matter-Power-Spectrum}. 

The upper and lower bounds of the integration have a great effect on the abundance. 
The minimum mass of the minihalo $M_{\rm min}$ should be determined by the Jeans scale
of IGM, which leads to
\begin{equation}
    M_{\rm min}
    = \frac{4\pi\rho_{m}(z)}{3}\left(\frac{5\pi k_{B}T_{\rm IGM}}{3G\mu\rho_{m}(z)}\right)^{3/2}
    \simeq 3.58\times10^{5}h^{-1}M_{\odot}\left(\frac{T_{\rm IGM}/K}{1+z}\right)^{3/2}.
\end{equation}
where $T_{\rm IGM}$ is the IGM temperature.
The IGM temperature around $z \sim 10$ is still unclear 
because of the uncertainties of the astrophysics. 
In this paper, 
we choose $T_{\rm IGM}=10\ {\rm K}, 100\ {\rm K}$ 
which avoid recent constraints on $T_{\rm IGM}$~\cite{Greig:2020hty,Greig:2020suk}.
\footnote
{
    The IGM temperature must be at least larger than the adiabatic temperature of the matter component.
    Because the matter temperature decreases adiabatically $\propto a^{-2}$ 
    after the decoupling from the radiation via the compton scattering around $z \simeq 150$~\cite{Madau:2003ee} without other heating,
    the adiabatic temperature is estimated as
    \[
    T_{\rm ad} \simeq \frac{2.73}{1+150}(1+z)^{2} \simeq 1.8\times10^{-2}(1+z)^{2}\ {\rm K},
    \]
    which is also excluded by recent observations around $z \sim 10$~\cite{Greig:2020hty,Greig:2020suk}.
}

The maximum mass $M_{\rm max}$ is determined by the condition 
\begin{align}
T_{\rm vir}\left(M\right)\apprle10^{4}\ {\rm K}.
\end{align}
Below this temperature, the star formation becomes inefficient 
because of the weakness of the metal-free gas cooling~\cite{Sheth:1999mn,Iliev:2002gj}.
The corresponding mini-halo mass is about $3 \times 10^7 M_{\odot}$.

The result of the calculation is shown in Fig.~\ref{fig:absorbers}.
All dashed lines (ALP without oscillons) are smaller than the CDM case
because the number of mini-halos is suppressed by homogeneous ULAP
as mentioned in Sec.~\ref{sec:Matter-Power-Spectrum}.
On the other hand, the solid lines  (ALP with oscillons) are larger than the dashed lines 
due to the enhanced number of mini-halos with $10^6M_\odot \lesssim M \lesssim 10^7M_\odot$.
Because the number of intersections is smaller than $1$ when $m \lesssim 10^{-21}\text{eV}$ 
even for ULAP oscillons with $T_{\rm IGM} = 10\ {\rm K}$,
it could be difficult to observe the difference of the 21cm absorption lines in this range.
The range becomes smaller 
when the IGM temperature is larger as shown in the lower figure of Fig.~\ref{fig:absorbers},
but this result still contains various uncertainties in $T_{\rm IGM}$ and the mini-halo profile.
Thus, we should wait for observational and theoretical progress for more precise estimation.


\section{Discussion}
\label{sec:discussion}

\begin{figure}
\centering{}
\includegraphics[width=1.0\textwidth]{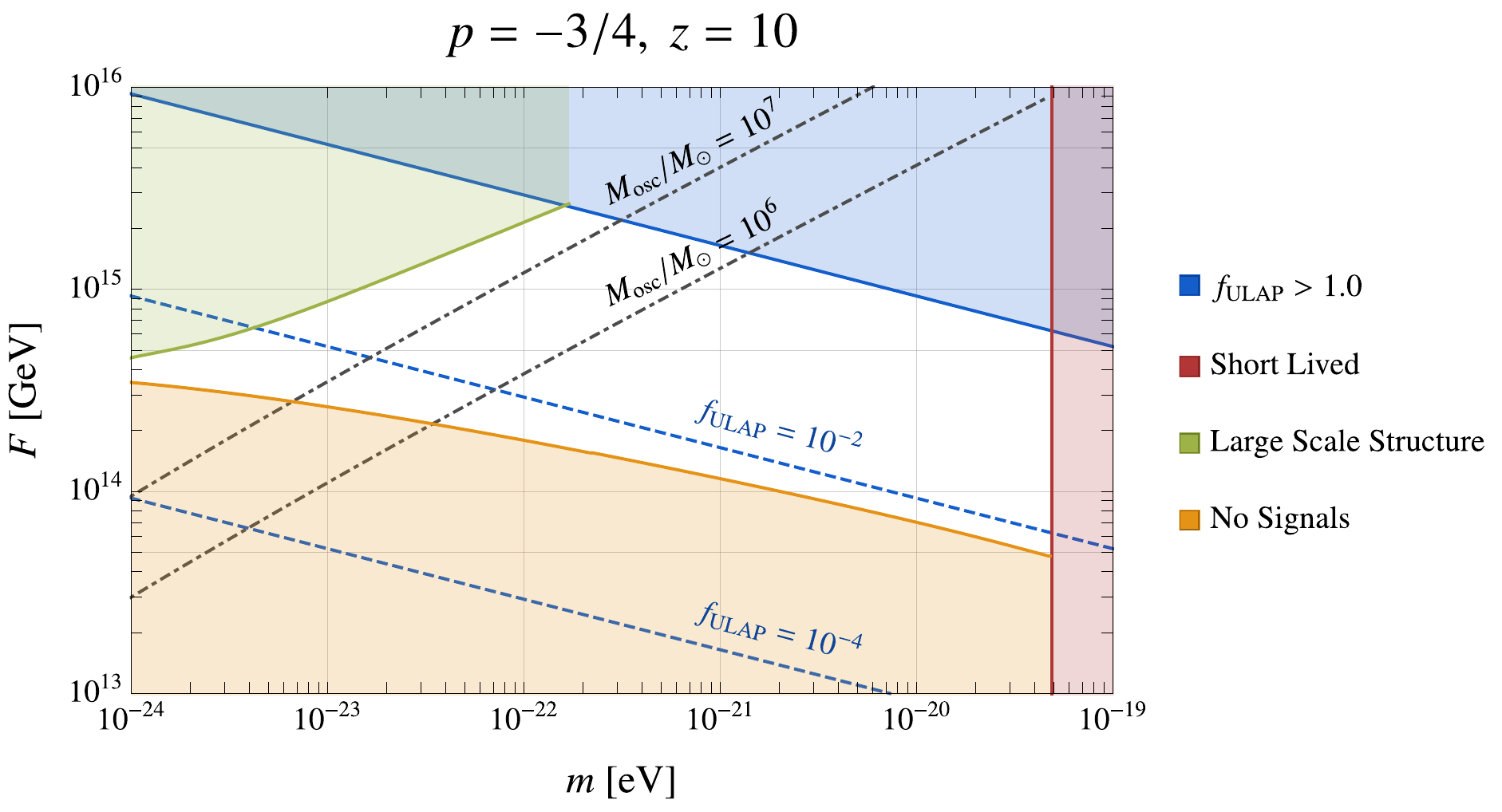}
\caption{
    The detectable parameter region by 21cm forest with ULAP oscillons.
    The horizontal and vertical axis show the ULAP mass and the ULAP decay constant, respectively.
    The blue region is excluded because the ULAP abundance $f_{\rm ULAP}$ is larger than $1$
    when the initial amplitude of ULAP is $\phi_i/F = 12\pi$.
    The red region shows the case that the lifetime of produced oscillons for $p = -3/4$ is smaller than the cosmic time at $z = 10$.
    The green region is excluded by the condition $P_{\rm osc}(1/{\rm Mpc}) > P_{\Lambda{\rm CDM}}(1/{\rm Mpc})$
    because the matter power spectrum on the large scale $\lesssim \mathcal{O}(1)\ {\rm Mpc}^{-1}$ is constrained by observations.
    The orange region is excluded by $P_{\rm osc}(k_{\rm cut}) < P_{\Lambda{\rm CDM}}(k_{\rm cut})$
    where $k_{\rm cut}$ is the cut-off wavenumber determined by the number of oscillon as mentioned in Sec.~\ref{sec:Matter-Power-Spectrum}.
    }
    \label{fig:detectability}
\end{figure}

\paragraph{Detectability}
We plotted the parameter region of ULAP that can be detectable by 21cm forest in Fig.~\ref{fig:detectability}.
We have considered the following constraints in the figure.
\begin{itemize}
    \item ULAP abundance: 
          Because the energy density at the oscillon formation is determined by the lattice simulation, 
          we can constrain the ULAP parameters by requiring $f_{\rm ULAP} < 1$, which excludes the blue region in Fig.~\ref{fig:detectability}. 
          Note that this constraint would be changed depending on the initial value of the ULAP.
          It would be stricter 
          when the initial amplitude becomes larger than our simulation value $\phi_i/F = 12\pi$, and vice versa.
    \item Oscillon lifetime:
          The produced oscillons must live up to the observation time ($z = 10$ in this paper)
          because the density fluctuation may be smeared out after the oscillon decay due to the self-radiation.
          This constraint for $p = -3/4$ is shown as the red region in Fig.~\ref{fig:detectability}.
    \item Observations of matter power spectrum:
          The matter power spectrum on the large scale $k \lesssim \mathcal{O}(1)\ {\rm Mpc^{-1}}$ 
          is precisely determined by many observations, 
          such as Planck~\cite{Aghanim:2018eyx}, DES~\cite{Abbott:2017wau}, and SDSS~\cite{2017AJ....154...28B, 2018ApJS..235...42A}.
          Thus, we constrain the ULAP parameters by the condition $P_{\rm osc}(1/{\rm Mpc}) > P_{\rm \Lambda CDM}(1/{\rm Mpc})$ as the green region in Fig.~\ref{fig:detectability}.
          We did not include the Lyman-$\alpha$ constraint discussed in Refs.~\cite{Irsic:2017yje, Kobayashi:2017jcf} here 
          because it is not obvious whether the produced oscillons affect the result of their simulations.
    \item The amplitude of the oscillon matter power spectrum:
          To detect the difference between the ULAP oscillon and the ordinary $\Lambda$CDM,
          the amplitude of the oscillon matter power spectrum must be at least larger than that of $\Lambda$CDM model.
          Thus, 
          the region $P_{\rm osc}(k_{\rm cut}) < P_{\Lambda{\rm CDM}}(k_{\rm cut})$ 
          is conservatively excluded
          where $k_{\rm cut}$ is the cut-off wavenumber mentioned in Sec.~\ref{sec:Matter-Power-Spectrum}.
          This constraint is shown as the orange region in Fig.~\ref{fig:detectability}
          and we find that ULAP is detectable if $f_{\rm ULAP} \gg 10^{-2}$.
\end{itemize}

\paragraph{Mini-halo profile}
When $p = -3/4$, the average oscillon mass at $z = 10$ is 
\begin{align}
    M_{\rm osc}
    \sim 6 \times 10^6 M_{\odot}\left(\frac{F}{10^{15}\ {\rm GeV}}\right)^2 \left(\frac{10^{-22}\ {\rm eV}}{m}\right).
\end{align}
while the interested mini-halo mass range is $10^6M_\odot \lesssim M \lesssim 10^7M_\odot$.
The contours of the constant oscillon mass $10^6 M_\odot$ and $10^7 M_\odot$ 
are plotted in Fig.~\ref{fig:detectability} as black dotted lines.
Between these lines,
the NFW profile may not describe the internal structure of the mini-halo well
because the mini-halo mass is almost the same as the produced oscillon mass.
In this case,
the mini-halo profile becomes more centered by oscillons,
which results in more absorption abundance near the mini-halo center.
However, it is unclear whether oscillons are disrupted by the gravitational force 
in the matter dominated era.
Thus, we used the NFW profile here and we will work on the gravitational stability of oscillons in future work.


\vspace{0.35cm}

Finally, we briefly comment on the existence of radio-loud sources in $z \gtrsim 10$ required for 21cm forest observations.
Recently, the radio loud sources around $z \sim 6$ 
with a flux $\gtrsim \mathcal{O}(10)\ {\rm mJy}$ sufficient for Square Kilometer Array (SKA) observations~\cite{Furlanetto:2006dt}
have been confirmed~\cite{Ba_ados_2018,Belladitta:2020zbp}
and a simple estimation indicates $10^4 \sim 10^5$ quasars around $z \simeq 10$ in the whole sky per redshift interval~\cite{Xu_2009,Shimabukuro:2019gzu,Shimabukuro:2020tbs}.
Besides, Population (Pop) III stars have been proposed to produce GRBs~\cite{10.1111/j.1745-3933.2009.00792.x,article,Meszaros:2010jw,Suwa:2010ze}, 
which could be unique sources in the high-redshift universe.
These possibilities support the importance of studies on the 21cm forest.


\section{Conclusion}
\label{sec:conclusion}

In this paper, 
we calculated the abundance of 21cm absorption lines 
when ULAP partially exists in the form of oscillons.
Because the structure on a scale determined by ULAP mass is significantly affected by oscillons,
the abundance of 21cm absorption lines is also changed.
We found that the matter power spectrum can be affected 
when the ULAP mass is $m \lesssim 10^{-19}\ {\rm eV}$ and the ULAP fraction is $f_{\rm ULAP} \gg 10^{-2}$.
This result is applicable to all ULAP models
which produce long-lived oscillons.

Unlike the previous researches,
because the Poisson-like power spectrum is cut off by the energy conservation on large scale,
we can focus on the phenomenologically interesting region $m \sim 10^{-22}\ {\rm eV}$ in this case.
Besides,
because the oscillons produce large fluctuations on a certain scale,
ULAP can be detectable
even if $f_{\rm ULAP}$ is smaller than $1$.

In this paper, 
we only focus on oscillons that survive at the observation time $z=10$.
However, the relativistic ULAP emitted by the complete decay of oscillons smears out the structure of the horizon scale
like warm dark matter,
which may give us another constraint on ULAP to consider.
This problem remains as future work.


\begin{acknowledgments}

We would like to thank Hayato Shimabukuro for very useful comments.
This work is supported by JSPS KAKENHI Grant Nos. 17H01131 (M.K.), 17K05434 (M.K.),
19H05810 (W.N.), 19J21974 (H.N.), and 19J12936 (E.S.),
World Premier International Research Center Initiative (WPI Initiative), MEXT, Japan, 
and Advanced Leading Graduate Course for Photon Science (H.N.).

\end{acknowledgments}


\bibliographystyle{JHEP}
\bibliography{refs}

\appendix

\section{Oscillon Matter Power Spectrum}
\label{sec:appnedix}

In this appendix,
we briefly introduce the analytical formula of the oscillon matter power spectrum at oscillon formation Eq.~(\ref{eq:amplitude_estimation}).
For simplicity, 
we will ignore the size of oscillons
and treat them as a point-like mass.
Thus, the derivation below is independent of the oscillon profile.

Let us consider a box with a comoving volume $V$
in which the number of oscillons with mass $M_i$ is represented by $N_i$.
When the positions of oscillons are not correlated, 
$N_i$ follows the Poisson distribution, i.e.
\begin{align}
    \Braket{(N_i-\braket{N_i})(N_j-\braket{N_j})}_N 
    = \braket{N_i}_N\delta_{ij},
\end{align}
where $\langle\rangle_N$ is statistical average over $N_i$.
Given that the density contrast is represented by
\begin{align}
    \delta  = \frac{\sum_i N_i M_i - \braket{\sum_i N_i M_i}_N}{\braket{\sum_i N_i M_i}_N},
\end{align}
and the physical number density of oscillons is 
\begin{align}
   n_{\rm osc} = \frac{\sum_i \braket{N_i}_N}{a^3V},
\end{align}
the oscillon power spectrum at $k = 2\pi V^{-1/3}$ is derived as
\begin{align}
   P_{\rm osc}(\bm{k})
  &= V \braket{\delta^2 }_N
\\&  =
    V \frac{1}{\braket{\sum_i N_i M_i}_N^2}
    \sum_{i,j} M_i M_j \Braket{(N_i-\braket{N_i})(N_j-\braket{N_j})}_N
\\&  =
    V \frac{1}{\braket{\sum_i N_i M_i}_N^2}
    \sum_{i} M_i^2 \braket{N_i}_N
\\&  =
    \frac{1}{n_{\rm osc}a^3} \frac{\Braket{M^2}}{\braket{M} ^2}
\end{align}
where we have defined the average of the oscillon mass and squared oscillon mass as
\begin{align}
    \braket{M} 
    \equiv  
    \frac{\sum_i  M_i\braket{N_i}_N}{\sum_i \braket{N_i}_N},
    \quad
    \Braket{M^2}
    \equiv  
    \frac{\sum_i M_i^2\braket{N_i}_N}{\sum_i \braket{N_i}_N}.
\end{align}
Note that the statistical average over $N_i$ is substituted by the simulation result in this paper.

In the above calculation, 
we have assumed that each oscillon number $N_i$ follows the Poisson distribution,
but the energy conservation constrains it
when the box size $V$ is larger than the scale of the energy transfer at oscillon formation.
Suppose the scale as $L_s$, 
the power spectrum is obtained with the suppression factor~\cite{Kawasaki:2020jnw} as
\begin{align}
  P_{\rm osc}(\bm{k})
  &=
  \frac{1}{n_{\rm osc}a^3} \frac{\Braket{M^2}}{\braket{M} ^2} K(kL_s),  \\
  K(x) 
  &= 
  \left[1-\left(\frac{2}{x}\right)^{2}\sin^{2}\left(\frac{x}{2}\right)\right],
\end{align}
which corresponds to Eq.~(\ref{eq:amplitude_estimation}) except for the overall factor related to the energy fraction.

\section{Simulation Setup}
\label{sec:setup}

\begingroup
\renewcommand{\arraystretch}{1.2}
\begin{table}[t]
    \centering
    \caption{
        Simulation parameters.
    }
    \vspace{0.3cm}
    \begin{tabular}{cc}
        \hline \hline
        $p$ & $-3/4$ \tabularnewline
        Box size  $L$ & $32$ \tabularnewline
        Grid size $N$ & $1024^3$ \tabularnewline
        Time          & $1-81$ \tabularnewline
        Time step     & $8 \times 10^{-3}$ \tabularnewline
        \hline \hline
    \end{tabular}
    \label{Ta:params}
\end{table}
\endgroup

In the simulation,
the units of the field, the conformal time, and the space, etc. are taken to be $F$ and $m^{-1}$, that is,
\begin{equation}
    \bar{\phi} \equiv \frac{\phi}{F},\ \
    \bar{\tau} \equiv m\tau,\ \
    \bar{x} \equiv mx,\ \ \dots\ {\rm etc}.
\end{equation}
where the overline denotes the dimensionless program variables and $\tau$ is the conformal time.

As the initial condition, 
we take the initial Hubble parameter as $H_i = 1/2t = m$
because ULAP starts to oscillate in the radiation dominated universe.
The initial scale factor is set to be unity $a_i = 1$, related to the conformal time as $a = \bar{\tau}$.
The initial field value and its derivative are set as
\begin{eqnarray}
    \bar{\phi}_i(\bm{x}) = 12\pi (1 + \zeta(\bm{x})),
    \quad
    \bar{\phi}_i(\bm{x})' = 0,
\end{eqnarray}
where dash represents the derivative with $\bar{\tau}$ and $\zeta(\bm x)$ is the initial noise defined by the scale-free power spectrum
\begin{align}
	\langle \xi_{\bm k}\xi_{\bm k'} \rangle
	= (2\pi)^3 \delta^3(\bm k-\bm k') \frac{2\pi^2}{k^3 } \mathcal P_\xi,
\end{align}
with a small constant $\mathcal P_\xi=2.1\times 10^{-9}$ as a reference value.
Other simulation parameters are shown in Table~\ref{Ta:params}.

We utilize our lattice simulation code used in Refs.~\cite{Ibe:2019vyo, Ibe:2019lzv, Kawasaki:2019czd},
in which the time evolution is calculated by the fourth-order symplectic integration scheme 
and the spatial derivatives are calculated by the fourth-order central difference scheme.
We impose the periodic boundary condition on the boundary.

\end{document}